\documentclass[twocolumn,showpacs,amsmath,amssymb,prl,german]{revtex4}

\usepackage{graphicx}   
\usepackage{dcolumn}    
\usepackage{bm}         
\usepackage{dsfont}     
\usepackage{subfigure,textcomp}
\newcommand{\rd}{{\rm d}}
\newcommand{\re}{{\rm e}}

\newcommand{\kB}{k_{\rm B}}

\renewcommand{\Im}{\rm Im}

\begin{document}

\title[Heat]{Radiative heat transfer between dielectric bodies}

\author{Svend-Age Biehs} 

\affiliation{Institut f\"ur Physik, Carl von Ossietzky Universit\"at,
    D-26111 Oldenburg, Germany}

\date{December 23, 2005}

\begin{abstract}
The recent development of a scanning thermal microscope (SThM) has led to measurements of radiative
heat transfer between a heated sensor and a cooled sample down to the nanometer range. 
This allows for comparision of the known theoretical
description of radiative heat transfer, which is based on fluctuating electrodynamics, with experiment. 
The theory itself is a macroscopic theory, which can be expected to break down at distances much smaller than 
$10^{-8} {\rm m}$. 
Against this background it seems to be reasonable to revisit the known macroscopic
theory of fluctuating electrodynamics and of radiative heat transfer.   
\end{abstract}

\pacs{44.40.+a, 05.40.-a, 03.50.De}

\maketitle


The development of a new kind of scanning thermal microscope (SThM) by the experimental 
Energy and Semiconductor Research group in Oldenburg has made possible measurements of radiative
heat transfer down to the nanometer range. A heated sensor tip incorporating a thermo-couple 
with a tip radius of $60 {\rm nm}$ scans a cooled sample surface, registering the thermo-voltage in the sensor.     
In order to assure that the results are not influenced by convection or adsorbates on the 
sample surface, the SThM works under ultra high vacuum conditions. In the future, scanning thermal microscopy 
might become a technique supplementing the well-known scanning tunneling and atomic force microscopy.
On the other hand, the measurements with the SThM allow for comparision of the  
existing theory of radiative heat transfer with the experiment.  
This shall be motivation enough for revisiting the theory of radiative heat transfer. 

%
%
\vspace{-0.3cm}
\section{Introduction}

The radiative heat transfer between macroscopic dielectric bodies is a well-understood
physical phenomenom and the energy transported per unit time and unit area can be 
expressed by the Kirchhoff-Planck law~\cite{L.D.Landau_E.M.Lifschitz_1979} in the stationary case, i.e., 
when the bodies are in local thermal equilibrium. But this is only true for distances $d$
which are large compared with the thermal wave length $\lambda_{{\rm th}}$ given 
by Wien's law, i.e., for $d \gg \lambda_{{\rm th}}$. 
For distances $d \ll \lambda_{{\rm th}}$ the radiative heat transfer increases
strongly due to so-called evanescent modes. 

\begin{figure}
  \centering
  \includegraphics[angle=0.0, height = 4.3cm]{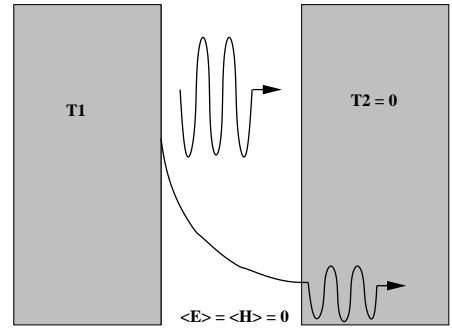}
  \caption{Sketch of the propagating and evanescent modes contributing to
           the radiative heat transfer between two semi-infinite bodies at different
           temperatures.}
  \label{Fig:SemiInfiniteBodies}
\end{figure}

For convenience we consider two semi-infinite bodies with a vacuum gap of distance $d$ 
between them (see Fig.~\ref{Fig:SemiInfiniteBodies}). This system is invariant under rotations and we chose the $z$-axis as
the axis of rotation. We further assume that the first body at $z < 0$ has a temperature 
$T_1 \ne 0$, whereas the second semi-infinite body at $z > d$ has zero temperature, i.e.,
$T_2 = 0$. Due to quantum and vacuum fluctuations of the atoms and electrons in the
first body there will be a fluctuating electromagnetical field inside that body.
Now, the boundary conditions of macroscopic electrodynamics require that the transversal
components of the electric and the magnetic field are continuous at the interface $z = 0$.
Therefore, there will also be a fluctuating electromagnetic field outside the semi-infinite
body produced by fluctuating internal sources. This field can be divided into a
propagating and an evanescent part. This becomes clear if we consider a plane wave solution
$\exp({\rm i} (k_z z - \omega t))$ in the vacuum gap propagating into the positive $z$-direction, 
with the modulus of the wave vector given by 
\begin{equation}
  k_z = \sqrt{\frac{\omega^2}{c^2} - k_\perp^2} =
           \begin{cases}
               \text{real}  &,  k_\perp < \frac{\omega}{c}, \\
               \text{complex}  &,   k_\perp >\frac{\omega}{c}.
           \end{cases}
\end{equation} 
Obviously, the modulus of the wave vector $k_z$ is purely real if the modulus of 
the transversal wave vector $k_\perp$ is smaller than the modulus of the vacuum wave
vector, $k_0 = \omega c^{-1}$. In this case we will have an oscillatory solution, i.e.,
propagating waves. On the other hand, $k_z$ is purely imaginary if $k_\perp > k_0$.
In that case the plane wave is exponentially damped in $z$-direction, i.e., the solution
is a so-called evanescent wave.

The characteristic wave length of the fluctuating field in the vacuum gap, generated in the semi-infinite
body at $z < 0$, is the thermal wave length $\lambda_{{\rm th}}$. It follows
that for distances $d \gg \lambda$, i.e., in the far-field region, the evanescent field 
can not reach into the second body and therefore only propagating modes can contribute 
to the radiative energy transfer. This energy transfer --- as mentioned above --- can then be 
described with the Kirchhoff-Planck law and is smaller than the energy transfer 
between two black bodies, which is given by the Stefan-Boltzmann law~\cite{L.D.Landau_E.M.Lifschitz_1979}
\begin{equation}
  S_{{\rm BB}} = \sigma (T_1^4 - T_2^4)
  \label{Eq:StefanBoltzmann}
\end{equation}
with $\sigma = 5.67\cdot10^{-8} {\rm W}{\rm m}^{-2}{\rm K}^{-4}$. On the other hand, for distances $d$ between 
the two bodies smaller than $\lambda_{{\rm th}}$, i.e., in the near-field region, 
the evanescent waves generated in the first body can reach into the second body and propagate there. 
This can be seen as some kind of photon tunneling through a vacuum barrier (see Fig.\ \ref{Fig:SemiInfiniteBodies}). 
Hence, in the near-field 
the propagating and evanescent fluctuating fields both will contribute to the radiative energy transfer. In that
case, the energy transfer can be several orders of magnitude larger than the black body value 
(\ref{Eq:StefanBoltzmann}), as we will see later. 

%
%
\section{Theoretical description}

\vspace{-0.3cm}
\subsection{Fluctuating fields}

The theoretical treatment of radiative heat transfer is usually based on Rytov's fluctuating
electrodynamics~\cite{S.M.Rytov_et_al_1989}. Within this framework the macroscopic Maxwell equations
are augmented by fluctuating source currents $\mathbf{j}$, describing the fluctuating 
sources of the electric and magnetic field $\mathbf{E}$ and $\mathbf{H}$ inside
the dielectric dissipative body. The individual 
frequency components of these source currents $\mathbf{j}(\mathbf{r},\omega)$ are 
considered as stochastic Gaussian variables. Within this treatment the Maxwell
equations become so-called stochastic Maxwell equations
\begin{align}
  \nabla\times\mathbf{E}(\mathbf{r},\omega) &= {\rm i} \omega \mu_0 \mathbf{H}(\mathbf{r},\omega), \nonumber \\
  \nabla\times\mathbf{H}(\mathbf{r},\omega) &= \mathbf{j}(\mathbf{r},\omega) 
                                               - {\rm i}\omega \epsilon(\omega) \mathbf{E}(\mathbf{r},\omega) 
  \label{Eq:StochasticMaxwell}
\end{align}
with the permeability of the vacuum $\mu_0$ and the permittivity of the body $\epsilon(\omega)$.
By writing the stochastic Maxwell equations in the form of eq.\ (\ref{Eq:StochasticMaxwell}) the dielectric body 
containing the source currents is already assumed to be non-magnetic, homogeneous, isotropic and local.

The electric and magnetic fields described by the stochastic Maxwell equations are classical
stochastic processes and can formally be expressed as
\begin{align}
  \mathbf{E}(\mathbf{r},\omega) &= {\rm i} \omega \mu_0 \int\!\!\rd^3 r'\,\mathds{G}^E(\mathbf{r,r'}) \mathbf{j}(\mathbf{r'},\omega), \nonumber \\
  \mathbf{H}(\mathbf{r},\omega) &= {\rm i} \omega \mu_0 \int\!\!\rd^3 r'\,\mathds{G}^H(\mathbf{r,r'}) \mathbf{j}(\mathbf{r'},\omega) 
  \label{Eq:StochasticFields}
\end{align} 
where the integral has to be taken over the source region, i.e., over the volume of the dielectric body
containing the source currents. The tensors 
$\mathds{G}^E$ and $\mathds{G}^H$ are the dyadic electric and the magnetic Green's function. The
electric dyadic Green's function is a solution of the Helmholtz wave equation
\begin{equation}
  \bigl(\nabla\times\nabla\times - k^2 \bigr) \mathds{G}^{E}(\mathbf{r,r'}) 
                                = \mathds{1} \delta(\mathbf{r - r'})
\end{equation}
with the wave vector $\mathbf{k}$ and the unit matrix $\mathds{1}$. By definition, the electric dyadic Green's 
function satisfies the boundary conditions of the electric field. Even though a similiar statement holds 
for the magnetic Green's function, it is more convenient to construct the dyadic magnetic Green's from
the electric Green's function by means of the relation
\begin{equation} 
  \mathds{G}^H (\mathbf{r,r'}) = - \frac{{\rm i}}{\omega \mu_0} \nabla \times \mathds{G}^{E}(\mathbf{r,r'}),
\end{equation}
which is a direct consequence of Maxwell's equations.

The reformulation of the electric and magnetic field $\mathbf{E}$ and $\mathbf{H}$ as an integral 
over the source currents in eqs.\ (\ref{Eq:StochasticFields}) makes clear that the fluctuating properties
of the fields are directly determined by the fluctuational properties of the source currents.
In fact, if $\langle \mathbf{j} (\mathbf{r},\omega) \rangle$ and 
$\langle \mathbf{j}(\mathbf{r},\omega) \mathbf{j}^*(\mathbf{r'},\omega') \rangle$ 
are known, where the angular brackets symbolise the ensemble average, the corresponding averages and
correlation functions of the fields can be evaluated. Therefore, it remains to determine the 
stochastic properties of the source currents.

From the definition of the source currents as stochastic Gaussian variables describing the thermal and
quantum fluctuations inside a dielectric dissipative body, it should be clear that the ensemble average
of the source currents vanishes. It follows with (\ref{Eq:StochasticFields}) that
\begin{equation}
  \langle \mathbf{E}(\mathbf{r},\omega) \rangle = \mathbf{0} \quad \text{and} \quad 
  \langle \mathbf{H} (\mathbf{r},\omega)\rangle = \mathbf{0}.
\end{equation}
However, the correlations and therewith the fluctuations of the source currents are determined by 
the properties of the dielectric dissipative body. This connection between dissipation and fluctuation
can be expressed in a general way by means of the fluctuation-dissipation theorem~\cite{L.D.Landau_E.M.Lifschitz_1979},
which can be applied to our problem~\cite{S.M.Rytov_et_al_1989}. For a homogeneous, isotropic and local
dielectric dissipative body this yields
\begin{equation}
  \begin{split}
    \langle j_\alpha(\mathbf{r},\omega) j_\beta^*(\mathbf{r'},\omega') \rangle & \\ 
                     & \hspace{-2cm} = 4 \pi \omega E(\omega, T) \epsilon''(\omega) \delta_{\alpha\beta} 
                        \delta(\mathbf{r-r'}) \delta(\omega - \omega')    
  \end{split}
\label{Eq:FDT}
\end{equation}
for the components of the source currents, where we have written $\epsilon = \epsilon' + {\rm i} \epsilon''$, and 
$E(\omega, T)$ is the Einstein function
\begin{equation}
  E(\omega, T) = \frac{\hbar \omega}{2} + \frac{\hbar \omega}{\re^{\hbar \omega \beta} - 1}
\end{equation}
with the usual inverse temperature $\beta = 1/\kB T$ of the body. The appearance of the
Einstein function is implied by the fluctuation-dissipation theorem, which indeed is
a quantum mechanical operator relation, but the fluctuating fields in the framework of fluctuating
electrodynamics are classical. This is a somewhat weak point in the theory, but it was recently 
shown~\cite{M.Janowicz_et_al_2003} that a rigorous quantum electrodynamical treatment for dissipative media
leads to the same correlation functions for the fields as fluctuating electrodynamics. Hence,
it seems to be justified to work within the framework of fluctuating electrodynamics, which should be 
classified as a semi-classical theory.  

Now, it is possible to evaluate the correlation functions of the fluctuating fields. Using eqs.\ 
(\ref{Eq:StochasticFields}) together with the fluctuation-dissipation theorem in eq.\ (\ref{Eq:FDT})
yields
\begin{equation}
  \begin{split}
    \langle E_\alpha (\mathbf{r},t) H_\beta(\mathbf{r},t) \rangle
            &= \int_0^\infty\!\!\!\rd\, \omega\, E(\omega, T) \frac{\omega \epsilon''(\omega)}{\pi} (\mu_0^2 \omega^2) \\
            &\quad\times\int\!\!\rd^3 r'\, \biggl( \mathds{G}^E {\mathds{G}^H}^\dagger \biggr)_{\alpha\beta} + {\rm c.c.}
  \end{split}
\label{Eq:CorrelationEH}
\end{equation}
where the $\dagger$-symbol denotes the adjoint dyadic and c.c. is an abbreviation for the complex conjugated term. 
Similiar equations can be derived for $\langle E_\alpha (\mathbf{r},t) E_\beta(\mathbf{r},t) \rangle$ and
 $\langle H_\alpha (\mathbf{r},t) H_\beta(\mathbf{r},t) \rangle$, respectively. 
The ensemble averages of the physical quantities --- the Poynting vector, the energy density and the stress tensor --- 
of a dielectric dissipative body can be evaluated with the help of these field correlation functions, if the
material properties, i.e., the permittivity $\epsilon(\omega)$, and the temperature of the body are known. Furthermore,
it is necessary to calculate the dyadic Green's functions for the given electrodynamical problem, in which
the geometry of the problem is incorporated.    

\vspace{-0.3cm}
\subsection{Radiative heat transfer between two slabs}

In principle, we are now able to calculate the radiative heat transfer between dielectric bodies of
arbitrary shape and volume as long these bodies are homogeneous, isotropic, and local dissipative dielectrics,
and can be described macroscopically. Unfortunately, the calculation of the dyadic Green's function
for the most geometries is rather complicated. Therefore, we will consider only the radiative heat transfer
between two bodies of the simplest possible geometry, i.e., two semi-infinite dielectric bodies separated
by a vacuum gap of width $d$ (see Fig.\ \ref{Fig:SemiInfiniteBodies}). 

The solution to this problem is well-known: the first derivation in the framework of Rytov's fluctuating
electrodynamics was given by Polder and van Hove~\cite{D.Polder_M.VanHove_1971}. Levin {\itshape et al.}~\cite{M.L.Levin_et_al_1980}
solved the problem with the surface impedance method of Leontovich, and Loomis and Maris~\cite{J.J._Loomis_H.J._Maris_1994} 
reinvestigated the problem some years ago. The radiative heat transfer between the two slabs can 
be stated as~\cite{D.Polder_M.VanHove_1971}   
\begin{equation}
  \begin{split}
    \langle S \rangle &= \frac{1}{4 \pi^2} \int_0^\infty\!\!\!\rd \omega\, \bigl[ E(\omega,T_1) - E(\omega,T_2)\bigr] \\
                      &\quad\times\int_0^\infty \!\!\! \rd k_\perp\, k_\perp \bigl\{ T_\parallel^{12} + T_\perp^{12} \bigr\}
  \end{split}
  \label{Eq:HeatTransfer}
\end{equation}
where $T^{12}_\parallel$ and $T_\perp^{12}$ are the 
transmission coefficients from the first body at $z < 0$ to the second body at $z > d$ for the TM- and TE-modes,
respectively. These transmission coefficients can be expressed by means of the usual Fresnel 
coefficients~\cite{L.D.Landau_E.M.Lifschitz_1974}, which gives for the propagating modes, i.e., for $k_\perp < k_0$,
\begin{equation}
  (T_\parallel^{12})^{\rm{prop}} = \frac{(1 - |r^{1}_\parallel|^2)(1 - |r^{2}_\parallel|^2)}{|1 - r^{1}_\parallel r^{2}_\parallel \rm{e}^{2 \rm{i} k_{z} d}|^2},
  \label{Eq:TransPr}  
\end{equation} 
and for the evanescent modes, i.e., for $k_\perp > k_0$,
\begin{equation}
  (T_\parallel^{12})^{\rm{ev}} = \frac{4 \Im(r^{1}_\parallel) \Im(r^{2}_\parallel) \rm{e}^{-2 \gamma d}}{|1 - r^{1}_\parallel r^{2}_\parallel \rm{e}^{- 2 \gamma d}|^2},
  \label{Eq:TransEv}
\end{equation} 
where the transmission coefficents for the TE-modes can be deduced from these relations by the 
substitution $\parallel \rightarrow \perp$.

The mean energy transfer per unit time and unit area between the two semi-infinite bodies at
different temperatures in eq.\ (\ref{Eq:HeatTransfer}) consists of two parts.
The temperature enters only into the first part, 
which is given by the difference of the Einstein functions evaluated at the coresponding
temperatures. From this difference it becomes clear that the vacuum term in the Einstein functions 
vanishes; there will be no vacuum contribution to the energy transfer. The second part is the integral
over the lateral wave vector $k_\perp$ which counts the modes contributing to the energy transfer. 
It can be shown that the above transmission coefficients are always smaller than $1$. Moreover, it
is possible to retrieve the energy transfer between two black bodies, which by definition have zero Fresnel
coefficients. It follows that the transmission coefficient for the evanescent modes (\ref{Eq:TransEv}) becomes zero and 
the transmission coefficient for the propagation modes (\ref{Eq:TransPr}) becomes~$1$. Inserting these transmission 
coefficients in eq.\ (\ref{Eq:HeatTransfer}) yields the Stefan-Bolzmann law stated in eq.\ (\ref{Eq:StefanBoltzmann}). 

From eq.\ (\ref{Eq:HeatTransfer}) together with the transmission coefficents it is possible to
study the radiative heat transfer between the two semi-infinite bodies in detail. At the moment,
we are only interested in the near-field limit of eq.\ (\ref{Eq:HeatTransfer}), i.e., in distances $k_0 d \ll 1$.
A Taylor expansion~\cite{M.L.Levin_et_al_1980} shows that in the near-field
\begin{equation}
  \langle S_\parallel \rangle \propto \frac{1}{d^2} \quad\text{and}\quad \langle S_\perp \rangle = {\rm const}.
  \label{Eq:TaylorExpansion}
\end{equation}
Therefore, the TM-mode part of the energy transfer dominates in the near-field region, 
$\langle S_\parallel \rangle \gg \langle S_\perp \rangle$, but it is not clear a priori at what distance
this domination begins. In fact, we will see by numerical calculations that the distance where the 
TM-mode contribution becomes dominant depends on the material properties of the two slabs. Furthermore,
from the near-field limit in eq.\ (\ref{Eq:TaylorExpansion}) two questions arise: i) Is a divergent radiative
energy transfer physically reasonable? ii) Is it possible to measure this power law? Or better, at what 
distance does the domination of the TM-mode part begin? 

The first question was discussed controversially~\cite{J.L.Pan_2000,J.P.Mulet_et_al_2001,J.L.Pan_2001}. 
But it should be clear that the description based on fluctuating electrodynamics
is still a macroscopic one, which means that at distances smaller
than $10^{-8}{\rm m}$ this theory is not valid and therefore does not lead to physically reasonable results.
In fact, the source currents in the fluctuation-dissipation theorem in eq.\ (\ref{Eq:FDT}) are delta-correlated 
with respect to $\mathbf{r - r'}$, but this can not be true at distances where the microscopic properties of
the materials make themselves felt. This delta-correlation of source currents seems to be the 
source of the divergent heat transfer at small distances, but in order to give a satisfactory answer to the 
question stated above a theory is needed that takes the finite correlations of source currents into account. 

\begin{figure}
  \centering
  \includegraphics[angle=0.0, height = 5.5cm]{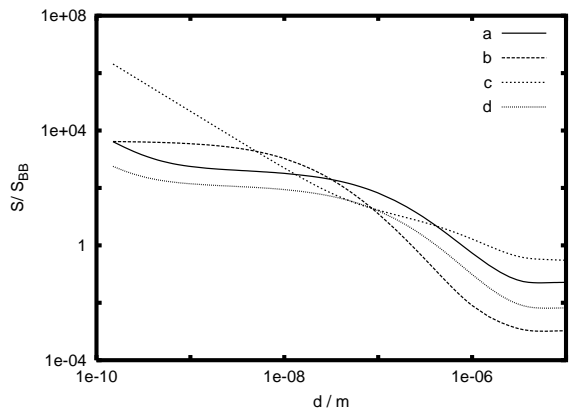}
  \caption{Numerical results of eq.\ (\ref{Eq:HeatTransfer}) for different Drude materials
           at different temperatures normalised to the black body value $S_{{\rm BB}}$ given
           by the Stefan-Boltzmann law in eq.\ (\ref{Eq:StefanBoltzmann}).}
  \label{Fig:NumericalResults}
\end{figure}

With this drawback of the macroscopic theory and the second question in mind we will now discuss some numerical
results of the heat transfer (\ref{Eq:HeatTransfer}) between two Drude materials~\cite{N.W.Ashcroft_N.D.Mermin_2001}. 
For convenience we set the temperature of the first body to $T_1 = 300 {\rm K}$ and the temperature of the 
second body to $T_2 = 0 {\rm K}$. Moreover, we use the same Drude permittivity for both bodies, i.e., 
$\epsilon_1 = \epsilon_2$.  

The numerical results for different plasma frequencies $\omega_p$ and relaxation times $\tau$
are given in Fig.\ \ref{Fig:NumericalResults}. For $a$, $b$, and $d$ we used relatively high relaxation
times and high plasma frequencies, so that $a$, $b$, and $d$ are plots for good conductors. On the 
other hand, for $c$ we used a relatively small relaxation time and small plasma frequency, which means that
$c$ is a plot for a bad conductor.

Obviously the divergency discussed before, which is associated with the 
TM-mode part of the radiative heat transfer $\langle S_\parallel \rangle$, can only be seen for $c$, i.e.,
for the bad conductor. There also is a divergency for good conductors, but it appears only for distances much smaller
than $10^{-9} {\rm m}$. Otherwise, in the region between $10^{-9} {\rm m}$ and $10^{-7} {\rm m}$
the radiative heat transfer for the good conductors $a$, $b$, and $c$ becomes constant. As discussed
before, such a behaviour can be attributed to the TE-mode part of the radiative heat transfer, $\langle S_\perp \rangle$. 
As an answer to the second question, it follows that for bad conductors the radiative heat transfer in 
the region between $10^{-9} {\rm m}$ and $10^{-7} {\rm m}$, which is accessible to measurements, is dominated by 
the TM-mode part of the radiative heat transfer, whereas for good conductors the TE-mode
part dominates. 

%
%
\section{Summary and outlook}

The radiative heat transfer between macroscopic dielectric bodies can be described 
by Rytov's theory of fluctuating electrodynamics. In the near-field the evanescent
modes give the main contribution to the radiative heat transfer. As the numerical 
results indicate, for good conductors the TE-mode part dominates the heat transfer
for experimentally accessible distances. In contrast, for bad conductors the 
TM-mode part dominates in that region and leads to a divergent radiative heat transfer.
This unsatisfactory feature of the theory can be traced back to the delta correlation in the correlation
function of the source currents. There is a need for a theory which takes a finite
correlation length of the source currents into account, and which should lead to a finite heat 
transfer at all distances. 

Experiments now being prepared in Oldenburg should answer some questions and thus  
provide a basis for such a theory.  Some of these questions are: At what distance does 
the macroscopic theory fail? What is the finite value of $\langle S \rangle$ for 
$d \rightarrow 0$? Is there an appropriate theory describing the sensor-sample geometry? 


\bibliography{biblio}
\bibliographystyle{prsty}

\end{document}